\documentclass[%
 reprint,
 nofootinbib,
 amsmath,amssymb,
 aps,
 pra,
]{revtex4-1}

\usepackage{graphicx}% Include figure files
\usepackage{dcolumn}% Align table columns on decimal point
\usepackage{bm}% bold math
\usepackage{bigints}% big integral signs
\usepackage{color}

\usepackage{booktabs,makecell}
\usepackage{hyphenat}

\begin{document}

\title{Comment on ``Retrieval of phase relation and emission profile of quantum cascade laser frequency combs''}% Force line breaks with \\
%\thanks{A footnote to the article title}%

\author{G\"unter Steinmeyer$^{1,2}$}
%\email{steinmey@mbi-berlin.de}
\affiliation{$^1$Max Born Institute for Nonlinear Optics and Short Pulse Spectroscopy, Max-Born-Stra{\ss}e 2a, 12489 Berlin, Germany}
\affiliation{$^2$Institut f\"ur Physik, Humboldt Universit\"at zu Berlin, Newtonstra{\ss}e 15, 12489 Berlin, Germany\\
email: steinmey@mbi-berlin.de}
\begin{abstract}
A recent publication on arXiv:1905.00668 suggests that the phase relationship within a frequency comb can be reconstructed from a heterodyne measurement using a reference comb. In principle, following this approach, such retrieval of optical phases appears possible, but should give rise to a temporally constant signal. Instead, arXiv:1905.00668 provides experimental evidence for a pronounced relative drift of the phases, which indicates that the modes of the comb are not coherently phase-locked. This temporal drift can be explained if the comb under test is not equidistant. This further means that the underlying temporal waveform is not repetitive and cannot be compressed into a coherent short pulse. This artifact severely hampers suggested applications of quantum cascade lasers in precision metrology. Moreover, the notion of designating multi-mode quantum cascade lasers as frequency combs appears questionable.
\end{abstract}

\maketitle

\noindent In Ref. \cite{arxiv}, Cappelli et al. suggest that the optical phases $\varphi_n$ of a frequency comb
\begin{equation}
E_{\rm FC}=\sum_n E_n \exp [ i (\omega_0 + n \omega_{\rm rep} t + \varphi_n ) ]
\end{equation}
can be extracted from heterodyning the said comb with a reference comb
\begin{equation}
E_{\rm ref}=\sum_n E'_n \exp [ i (\omega'_0 + n \omega'_{\rm rep}) t].
\end{equation}
For simplicity, we assume a flat spectral phase of the reference comb. Here $\omega_0$ and $\omega'_0$ designate the central frequencies (or carrier-envelope frequencies) of the combs. $\omega_{\rm rep}$ and $\omega'_{\rm rep}$ determine the respective intermode beat frequencies. In the following, we additionally set all amplitude factors to unity, i.e., $E'_n=E_n=1$ for simplicity.
The definition of Eq.~(1) in \cite{arxiv} already raises two questions. First, the electric field in that equation is not real-valued. Moreover, it is assumed that $\omega_{\rm rep}$ of the unknown comb is a constant with respect to frequency. Given that a quantum cascade laser (QCL) does not include any saturable absorption effect, it appears questionable whether a coherent phase-lock between the modes can be established and whether the blind assumption of an equidistant comb holds. In fact, the so-called cold-cavity modes of a multi-mode laser are known to be not equidistant due to chromatic dispersion of the intracavity chromatic dispersion. To avoid some of the shortcomings in \cite{arxiv}, a more generally applicable picture results from the definition
\begin{equation}
E_{\rm FC}=\sum_n a_n = \sum_n \cos [(\omega_{\rm CE} + n \omega_{\rm rep} + \epsilon_n) t + \varphi_n ]
\end{equation}
for the comb under test. Here an additional parameter $\epsilon_n$ is introduced to account for possible small deviations from equidistance of the comb.
One then defines
\begin{equation}
E_{\rm ref}=\sum_n b_n = \sum_n \cos [n \omega'_{\rm rep} t]
\end{equation}
for the reference comb. Heterodyning these two combs on a photo detector gives rise to the heterodyne signal
\begin{equation}
S=|E_{\rm FC}+E_{\rm ref}|^2 = \left( \sum_n a_n + \sum_m b_m \right)^2.
\end{equation}
Out of this heterodyne signal, one can select a particular beat frequency signal
\begin{equation}
S_{\rm beat}(n,m) = \cos [m \omega'_{\rm rep} t] \cos [(\omega_{\rm CE} + n \omega_{\rm rep} + \epsilon_n) t + \varphi_n ].
\end{equation}
Using the trigonometric identity $2 \cos a \cos b = \cos (a-b) + \cos (a+b)$, one isolates the rf signal as the difference frequency term
\begin{equation}
S_{\rm diff}(n,m) = \cos [ (\omega_{\rm CE} + n \omega_{\rm rep} - m \omega'_{\rm rep} + \epsilon_n) t + \varphi_n ].
\end{equation}
Ref.~\cite{arxiv} now suggests that we have phase-coherent access to the intermode beats and can exploit this for generating a phase-coherent reference signal at frequency
\begin{equation}
 \omega_{\rm ref} = n \omega_{\rm rep} - m \omega'_{\rm rep},
\end{equation}
which give us access to extracting the in-phase signal
\begin{equation}
S_{\rm I}(n) = \int\limits_{t_0}^{t_0 + T_{\rm int}} \cos(\omega_{\rm ref} t) S_{\rm diff}(n,m) {\rm d}t
\end{equation}
via lockin detection with integration time $T_{\rm int}$. In addition, we can retrieve the quadrature signal
\begin{equation}
S_{\rm Q}(n) = \int\limits_{t_0}^{t_0 + T_{\rm int}} \sin(\omega_{\rm ref} t) S_{\rm diff}(n,m) {\rm d}t.
\end{equation}
Using the same trigonometric relation as above we remove the rapidly oscillating components and simplify
\begin{eqnarray}
S_{\rm I}(n) & =  & \int\limits_{t_0}^{t_0 + T_{\rm int}} \cos [ (\omega_{\rm CE} + \epsilon_n) t + \varphi_n] {\rm d}t \label{res1} \\
S_{\rm Q}(n) & =  & \int\limits_{t_0}^{t_0 + T_{\rm int}} \sin [ (\omega_{\rm CE} + \epsilon_n) t + \varphi_n] {\rm d}t. \label{res2}
\end{eqnarray}
In fact, assuming $\omega_{\rm CE}=\epsilon_n=0$, the integration becomes trivial, and $\varphi_n$ can be extracted by computing $\arctan[S_{\rm I}(n)/S_{\rm Q}(n)]$. In the latter case, there is no remaining time dependence in the integrand, and repeated measurements of the phase should always yield the identical result. However, none of the measurements in \cite{arxiv} shows the behavior predicted by Eqs.~(\ref{res1}) and (\ref{res2}). Instead, all longterm measurements in \cite{arxiv} show pronounced temporal phase drifts in between the measurements. These drifts contradict the authors' initial assumption that the $\varphi_n$ of their comb be constant in time and only fluctuations $\delta\varphi_n \ll \pi$ be allowed. 

For example, the measurement in Fig.~2(d) of \cite{arxiv} shows irregular fluctuations $\delta\varphi_n \ll \pi$ over a range of 1300 degrees. As this measurement involved two combs with perfect equidistance such fluctuations can only be explained if the phase coherence was at least occasionally corrupted in the signal processing chain. As this happened repeatedly it severely questions conclusions drawn from the longterm measurements.

\begin{figure}[t!]
\centering
\includegraphics[width=0.7\linewidth]{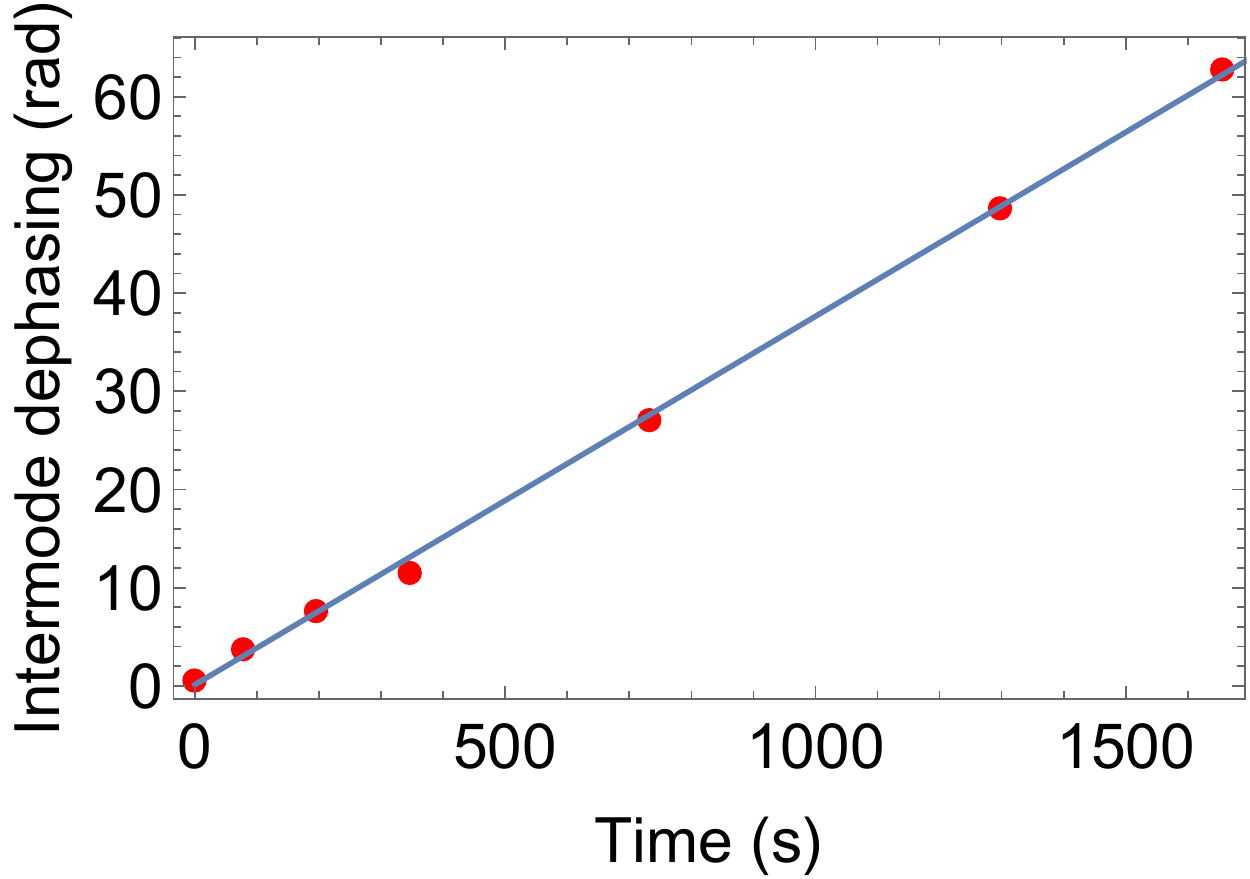}
\caption{Mode-to-mode phase slippage vs. measurement time. Data has been extracted from Fig.~4(d) of \cite{arxiv}. Phase unwrapping was applied according to the measured drift during the first 77\,s. Given the rather sparse phase sampling, this reconstruction may grossly underestimate the decoherence effects in the QCL.}
\label{fig}
\end{figure}

In contrast, the heterodyne measurement between a QCL comb and a pulsed fs comb in Fig.~4(d) of \cite{arxiv} shows a much more systematic and continuous variation of the phase slope. Looking at the development of the beat note during the first $77\,$s, the beat tone at 10\,MHz drifts by $8 \pi$ whereas the phase of the beat tone at 40\,MHz seems to be nearly constant over time. Extracting the data from \cite{arxiv}, one can attempt to reconstruct the mode-to-mode dephasing as a function of time, see Fig.~\ref{fig}. Given that the phase was only sampled sparsely at time intervals larger than one minute, this reconstruction may grossly underestimate the decoherence effects inside the laser. This behavior cannot simply be explained by imperfections in the carrier envelope frequency compensation or noise-induced errors in the unwrapping procedure, and this trend systematically continues over subsequent measurements of the heterodyne beats. This violates the stated precondition for a frequency comb. In fact, the most likely explanation appears to be a deviation from equidistance, i.e., $\epsilon_n$ is a non-trivial function of $n$.

The measurements in Ref.~\cite{arxiv} therefore appear to be the first clear evidence for long-suspected deviations from equidistance in QCL frequency combs. As the phase data in \cite{arxiv} have only been sparsely sampled in time, it is impossible to conclude from the published data how large the deviations from equidistance actually are. The only safe conclusion appears to be that the mode-to-mode deviation from equidistance is on the order of a hertz or smaller. Given that QCL combs often encompass several hundred modes, this means that deviations from equidistance may readily accumulate to tens or hundreds of hertz over the entire comb. Compared to typical intermode beat frequencies of several GHz, the deviation from equidistance is still a small number that has obviously been overlooked in previous measurements. Nevertheless, such deviations seem to rule out the usefulness of QCL combs in precision frequency metrology.

On the other hand, the non-equidistance of QCL combs does not appear overly surprising. QCLs exhibit an extremely short upperstate lifetime in the few-picosecond or even femtosecond range, which has to be compared to typical cavity roundtrip times of more than 100\,ps. Given that there is no established saturable absorption mechanism inside the QCL cavity, proven mode-locking theories \cite{Haus1,Haus2} clearly rule out the persistence of a repetitive waveform. It therefore appears that QCLs are simply multi-mode continuous-wave lasers. As such, they may still be useful, e.g., for dual-comb spectroscopy with kHz spectral resolution. Nevertheless, it appears advisable that the QCL community adopt to a more realistic view of the limitations of their so-called QCL frequency combs.

\end{document}